\documentclass{aastex}
\slugcomment{{\sc Accepted to ApJ:} January, 2014} 

\usepackage{natbib}
\usepackage{amsmath}
\usepackage{amssymb}
\usepackage{color}
\usepackage{ulem}

\def\tev{TeV J2032+4130}
\def\vtev{VER J2031+415}

\def\spec{\,TeV$^{-1}$\,cm$^{-2}$\,s$^{-1}$}

\usepackage{lineno}
\modulolinenumbers[5]

\begin{document}

\singlespace

\title{Observations of the unidentified gamma-ray source \tev\ by VERITAS}

\author{
E.~Aliu\altaffilmark{1},
T.~Aune\altaffilmark{2},
B.~Behera\altaffilmark{3},
M.~Beilicke\altaffilmark{4},
W.~Benbow\altaffilmark{5},
K.~Berger\altaffilmark{6},
R.~Bird\altaffilmark{7},
J.~H.~Buckley\altaffilmark{4},
V.~Bugaev\altaffilmark{4},
J.~V~Cardenzana\altaffilmark{8},
M.~Cerruti\altaffilmark{5},
X.~Chen\altaffilmark{9,3},
L.~Ciupik\altaffilmark{10},
M.~P.~Connolly\altaffilmark{11},
W.~Cui\altaffilmark{12},
C.~Duke\altaffilmark{13},
J.~Dumm\altaffilmark{14},
M.~Errando\altaffilmark{1},
A.~Falcone\altaffilmark{15},
S.~Federici\altaffilmark{3,9},
Q.~Feng\altaffilmark{12},
J.~P.~Finley\altaffilmark{12},
P.~Fortin\altaffilmark{5},
L.~Fortson\altaffilmark{14},
A.~Furniss\altaffilmark{16},
N.~Galante\altaffilmark{5},
G.~H.~Gillanders\altaffilmark{11},
S.~Griffin\altaffilmark{17},
S.~T.~Griffiths\altaffilmark{18},
J.~Grube\altaffilmark{10},
G.~Gyuk\altaffilmark{10},
D.~Hanna\altaffilmark{17},
J.~Holder\altaffilmark{6},
G.~Hughes\altaffilmark{3}  \footnote{gareth.hughes@desy.de},
T.~B.~Humensky\altaffilmark{19},
P.~Kaaret\altaffilmark{18},
O.~Kargaltsev\altaffilmark{32},
M.~Kertzman\altaffilmark{20},
Y.~Khassen\altaffilmark{7},
D.~Kieda\altaffilmark{21},
H.~Krawczynski\altaffilmark{4},
M.~J.~Lang\altaffilmark{11},
A.~S~Madhavan\altaffilmark{8},
G.~Maier\altaffilmark{3},
P.~Majumdar\altaffilmark{2,22}  \footnote{pratik.majumdar@saha.ac.in},
A.~McCann\altaffilmark{23},
P.~Moriarty\altaffilmark{24},
R.~Mukherjee\altaffilmark{1},
D.~Nieto\altaffilmark{1},
A.~O'Faol\'{a}in de Bhr\'{o}ithe\altaffilmark{7},
R.~A.~Ong\altaffilmark{2},
A.~N.~Otte\altaffilmark{25},
D.~Pandel\altaffilmark{26},
J.~S.~Perkins\altaffilmark{27},
M.~Pohl\altaffilmark{9,3},
A.~Popkow\altaffilmark{2},
H.~Prokoph\altaffilmark{3},
J.~Quinn\altaffilmark{7},
K.~Ragan\altaffilmark{17},
J.~Rajotte\altaffilmark{17},
L.~C.~Reyes\altaffilmark{28},
P.~T.~Reynolds\altaffilmark{29},
G.~T.~Richards\altaffilmark{25},
E.~Roache\altaffilmark{5},
G.~H.~Sembroski\altaffilmark{12},
C.~Skole\altaffilmark{3},
D.~Staszak\altaffilmark{17},
I.~Telezhinsky\altaffilmark{9,3},
M.~Theiling\altaffilmark{12},
J.~V.~Tucci\altaffilmark{12},
J.~Tyler\altaffilmark{17},
A.~Varlotta\altaffilmark{12},
S.~Vincent\altaffilmark{3},
S.~P.~Wakely\altaffilmark{30},
T.~C.~Weekes\altaffilmark{5},
A.~Weinstein\altaffilmark{8},
R.~Welsing\altaffilmark{3},
D.~A.~Williams\altaffilmark{16},
B.~Zitzer\altaffilmark{31}
}

\altaffiltext{1}{Department of Physics and Astronomy, Barnard College, Columbia University, NY 10027, USA}
\altaffiltext{2}{Department of Physics and Astronomy, University of California, Los Angeles, CA 90095, USA}
\altaffiltext{3}{DESY, Platanenallee 6, 15738 Zeuthen, Germany}
\altaffiltext{4}{Department of Physics, Washington University, St. Louis, MO 63130, USA}
\altaffiltext{5}{Fred Lawrence Whipple Observatory, Harvard-Smithsonian Center for Astrophysics, Amado, AZ 85645, USA}
\altaffiltext{6}{Department of Physics and Astronomy and the Bartol Research Institute, University of Delaware, Newark, DE 19716, USA}
\altaffiltext{7}{School of Physics, University College Dublin, Belfield, Dublin 4, Ireland}
\altaffiltext{8}{Department of Physics and Astronomy, Iowa State University, Ames, IA 50011, USA}
\altaffiltext{9}{Institute of Physics and Astronomy, University of Potsdam, 14476 Potsdam-Golm, Germany}
\altaffiltext{10}{Astronomy Department, Adler Planetarium and Astronomy Museum, Chicago, IL 60605, USA}
\altaffiltext{11}{School of Physics, National University of Ireland Galway, University Road, Galway, Ireland}
\altaffiltext{12}{Department of Physics, Purdue University, West Lafayette, IN 47907, USA }
\altaffiltext{13}{Department of Physics, Grinnell College, Grinnell, IA 50112-1690, USA}
\altaffiltext{14}{School of Physics and Astronomy, University of Minnesota, Minneapolis, MN 55455, USA}
\altaffiltext{15}{Department of Astronomy and Astrophysics, 525 Davey Lab, Pennsylvania State University, University Park, PA 16802, USA}
\altaffiltext{16}{Santa Cruz Institute for Particle Physics and Department of Physics, University of California, Santa Cruz, CA 95064, USA}
\altaffiltext{17}{Physics Department, McGill University, Montreal, QC H3A 2T8, Canada}
\altaffiltext{18}{Department of Physics and Astronomy, University of Iowa, Van Allen Hall, Iowa City, IA 52242, USA}
\altaffiltext{19}{Physics Department, Columbia University, New York, NY 10027, USA}
\altaffiltext{20}{Department of Physics and Astronomy, DePauw University, Greencastle, IN 46135-0037, USA}
\altaffiltext{21}{Department of Physics and Astronomy, University of Utah, Salt Lake City, UT 84112, USA}
\altaffiltext{22}{Saha Institute of Nuclear Physics, Kolkata 700064, India}
\altaffiltext{23}{Kavli Institute for Cosmological Physics, University of Chicago, Chicago, IL 60637, USA}
\altaffiltext{24}{Department of Life and Physical Sciences, Galway-Mayo Institute of Technology, Dublin Road, Galway, Ireland}
\altaffiltext{25}{School of Physics and Center for Relativistic Astrophysics, Georgia Institute of Technology, 837 State Street NW, Atlanta, GA 30332-0430}
\altaffiltext{26}{Department of Physics, Grand Valley State University, Allendale, MI 49401, USA}
\altaffiltext{27}{N.A.S.A./Goddard Space-Flight Center, Code 661, Greenbelt, MD 20771, USA}
\altaffiltext{28}{Physics Department, California Polytechnic State University, San Luis Obispo, CA 94307, USA}
\altaffiltext{29}{Department of Applied Physics and Instrumentation, Cork Institute of Technology, Bishopstown, Cork, Ireland}
\altaffiltext{30}{Enrico Fermi Institute, University of Chicago, Chicago, IL 60637, USA}
\altaffiltext{31}{Argonne National Laboratory, 9700 S. Cass Avenue, Argonne, IL 60439, USA}
\altaffiltext{32}{Department of Physics, The George Washington University, Washington, DC 20052, USA}

\maketitle

\section*{Abstract}

\tev\ was the first unidentified source discovered at very high energies (VHE; E $>$ 100 GeV), with
no obvious counterpart in any other wavelength. It is also the first extended source to be observed in
VHE gamma rays. Following its discovery, intensive observational campaigns have been
carried out in all wavelengths in order to understand the nature of the object, which have met with limited success.
We report here on a deep observation of \tev\, based on 48.2 hours of data taken from 2009 to 2012
by the VERITAS (Very Energetic Radiation Imaging Telescope Array System) experiment. The source is detected
at 8.7 standard deviations ($\sigma$) and is found to be
extended and asymmetric with a width of 9.5$^{\prime}$$\pm$1.2$^{\prime}$ along the major axis and 
4.0$^{\prime}$$\pm$0.5$^{\prime}$ along the minor axis. The spectrum is well
described by a differential power law with an index of 2.10 $\pm$ 0.14$_{stat}$ $\pm$ 0.21$_{sys}$ and a normalization of
(9.5 $\pm$ 1.6$_{stat}$ $\pm$ 2.2$_{sys}$) $\times$ 10$^{-13}$\spec\ at 1 TeV.
We interpret these results in the context of multiwavelength scenarios which particularly favor the pulsar wind nebula
(PWN) interpretation.

\medskip{\bf Keywords:} gamma-ray sources: individual: (TeV J2032+4130 = VER J2031+415), pulsars: individual (PSR J2032+4127)

\section{Introduction}

The Cygnus X complex is one of the brightest areas of the sky in all wavelengths and is host to a large 
number of sources and source types. These include active star forming regions, pulsars, and supernova 
remnants (SNRs). Objects connected to this region include the star association Cygnus OB2, the microquasar 
Cygnus X-3, the SNR G78.2+2.1, and \tev\, making it a natural laboratory for the study of cosmic ray 
acceleration. \tev\ was discovered serendipitously by the HEGRA imaging atmospheric Cherenkov telescope 
(IACT) system \citep{2002A&A...393L..37A,2005A&A...431..197A} during observations made in the years 
1999--2001. It was the first TeV gamma-ray detection to have no obvious counterpart at any other 
wavelength and was also the first extended source to be discovered in the VHE
range. Analysis of combined HEGRA data from 1999--2002 gave a final position for the extended VHE 
source of R.A. = 20$^{h}$31$^{m}$57.0$^{s}$ $\pm$ 6.2$^{\prime}$$_{stat}$ $\pm$ 
13.7$^{\prime\prime}$$_{sys}$, Decl. = +41$^\circ$29$^{\prime}$57$^{\prime\prime}$ $\pm$ 
1.1$^{\prime}_{stat}$ $\pm$ 1.0$^{\prime}_{sys}$ (excess events center of gravity), and a Gaussian 
standard deviation width of 
$\sigma$ = 6.2$^{\prime}$ $\pm$ 1.2$^{\prime}_{stat}$ $\pm$ 0.9$^{\prime}_{sys}$ 
\citep{2005A&A...431..197A}. HEGRA reported an integral gamma-ray flux above 1 TeV of  
 (6.9 $\pm$ 1.8) $\times$ 10$^{-13}$ cm$^{-2}$ s$^{-1}$ with spectral index of 1.9 $\pm$ 0.3.

Following the discovery, an archival search of the Whipple 10m telescope data was carried out that showed
evidence of a source consistent with \tev\ \citep{2004A&A...423..415}. However, it is worth
noting that the peak emission in the Whipple data had an offset of $\sim$ 3.6$^{\prime}$ with respect to the
HEGRA position. The gamma-ray flux measured by Whipple was 12\% of the Crab nebula flux above 400 GeV.
Based on observations carried out by Whipple in 2003-05, and assuming a spectral shape the same as that 
of the Crab nebula, it was later reported to be ~8\% of the Crab nebula flux~\citep{2007ApJ...658.1062K}.
The MAGIC collaboration, too, has reported a deep exposure of this region \citep{2008ApJ...675L..25A}.
The MAGIC collaboration has also found the source to be extended, with an integral flux 
and spectral index comparable to that measured by HEGRA.
These measurements have been extended to even higher energies by the air shower array detectors Milagro and ARGO
\citep{2012ApJ...753..159A,2013arXiv1311.3376T}.
Summaries of the positions and morphologies of the
results discussed can be found in Table \ref{tab:positions}.

Since the discovery of \tev, several observations of the region have been made by X-ray telescopes 
including $Chandra$ and $XMM$-$Newton$, which operate in the energy range of 0.1 -- 10 keV and 0.2 -- 12 keV respectively. 
Multiple ($\sim$~20) point sources were detected in a 5 ksec $Chandra$ 
observation \citep{2003ApJ...597..494B}. A deep follow-up 50 ksec observation yielded 240 X-ray sources within 
the same field of view \citep{2006ApJ...643..238B}. A $\sim$ 50~ksec $XMM$-$Newton$ exposure was also 
obtained \citep{2007A&A...469L..17H}. After the known X-ray sources were subtracted, \citet{2007A&A...469L..17H} 
reported an extended X-ray emission region with a FWHM of $\sim$12 arcmins. An analysis of the $Chandra$
data also showed the presence of diffuse X-ray emission; however, low photon statistics did not allow for 
a detailed study of the spectrum. \citet{2003ApJ...589..487M} carried out optical observations of 
several of the brightest X-ray sources and found that most of these were either O stars or foreground 
late-type stars.

Observations were also made by $Suzaku$~\citep{2011PASJ...63S.873M} in the energy range 2-10 keV. 
The authors found two structures within the 
TeV gamma-ray emission region. After estimating the contribution from the point sources identified by 
$Chandra$, 
the X-ray spectra of the 
diffuse components were extracted. The diffuse X-ray spectrum was best-fit with a power-law with a photon 
index of $\sim$~2.

Radio observations of the region have been made using the Giant Metrewave Radio Telescope 
\citep{2007ApJ...654L.135P}, yielding several radio sources. At least three of the sources were reported 
to be non-thermal along with an extended non-thermal diffuse emission.

A previously unknown gamma-ray pulsar, PSR J2032+4127, with a pulse period of 142 ms, was discovered in a blind search by 
$Fermi$-Large Area Telescope (LAT), located 0.07$^{\circ}$ from the center of the HEGRA detection~\citep{2009ApJ...705....1C}. 
Subsequent radio measurements made by the Green Bank Telescope \citep{2009ApJ...705....1C} localized the 
position to within a few arcseconds. A characteristic age of 0.11 Myr and a spin-down power of 2.7 
$\times$ 10$^{35}$ erg s$^{-1}$ were derived \citep{2009ApJ...705....1C}. A dispersion measure of 
114.8$\pm$1.0 pc cm$^{-3}$ resulted in a distance of 3.6 kpc when standard models for dispersion in the Milky 
Way were applied \citep{2002astro.ph..7156C}.  However, based on the pulsar's gamma-ray luminosity, a 
revised estimate of 1.7 kpc was suggested \citep{2009ApJ...705....1C}, 
which would place it at the same distance as the Cygnus OB2 star-forming region. The discovery of the pulsar has led 
several authors to establish a connection between \tev\ and PSR~J2032+4127 (for more details see \citet{2009RAA.....9..841C}).
This argument has been strengthened by the detection of the X-ray emission that is spatially coincident with \tev.  
However arguments based on spatial coincidence alone can be suspect due to the fact that the morphology of the 
source
can be different at different wavelengths. In order to understand the nature of the emission, 
it is important to study the
morphology of the source with a detector with an improved sensitivity and angular resolution. 
It is clear from the above discussions that despite several attempts to unravel the nature of \tev\ since its
discovery, the source along with its position, flux and morphology remains a mystery. 
This prompted VERITAS to perform a deep observation of this very interesting region of the Galaxy
with the aim of better understanding the morphology and source position.

The paper is split into several parts: Section 2 describes the VERITAS experiment and observations made on \tev. 
Section 3 describes the results obtained, and Section 4 presents a $Fermi$-LAT analysis of the region of \tev.  
Finally we discuss the implications of our observations on the source in the context of multiwavelength 
observations.

\begin{small}
\begin{table}
\begin{center}
\begin{tabular}{c|c|c|c|c|c}
  \hline
    Experiment & R.A. & Error in R.A.& Decl. & Error in Decl.  & Reported Extension \\
    & & $Stat(Sys)$ & & $Stat(Sys)$ & $\pm Stat(Sys)$ \\
    && (arcmin) && (arcmin) &(arcmin) \\
  \hline
  HEGRA    & 20$^{h}$31$^{m}$57$^{s}$   & 6.2(13.7) & +41$^{\circ}$29$^{\prime}$57$^{\prime \prime}$ & 1.1(1.0) & 6.2$\pm$1.2(0.9) \\
  Whipple  & 20$^{h}$32$^{m}$27$^{s}$   & 21(23)    & +41$^{\circ}$39$^{\prime}$17$^{\prime \prime}$ & 5(6)     & $<$6.0 \\
  MAGIC    & 20$^{h}$32$^{m}$20$^{s}$   & 11(11)    & +41$^{\circ}$30$^{\prime}$36$^{\prime \prime}$ & 1.2(1.8) & 5.0$\pm$1.7(0.6) \\
  Milagro  & 20$^{h}$28$^{m}$43.2$^{s}$ & 25        & +41$^{\circ}$07$^{\prime}$48$^{\prime \prime}$ & 16       & 66 \\
  ARGO     & 20$^{h}$32$^{m}$24.0$^{s}$ & -         & +41$^{\circ}$45$^{\prime}$00$^{\prime \prime}$ & -        & 12$^{+24}_{-12}$ \\
  \hline
\end{tabular}
\end{center}
\caption{Summary of VHE detections, measured positions and extensions for each experiment prior to VERITAS observations. 
Also shown are the reported right accessions and 
declinations (J2000), their corresponding errors and measured extensions of the source. 
Systematic errors are shown in parentheses.}
\label{tab:positions}
\end{table}
\end{small}

\section{VERITAS Observations}

The VERITAS array of IACTs is located at the Fred Lawrence Whipple observatory in southern Arizona (1.3 km 
a.s.l., N 31$^\circ$40$^{\prime}$, W 110$^\circ$57$^{\prime}$). It consists of four Davies-Cotton 
telescopes \citep{1964sew6.conf..336C}.  Full array operations began in September 2007. Each telescope has 
a focal length and a dish diameter of 12 meters and comprises 345 tessellated identical hexagonal mirror facets 
for a total effective mirror area of 106~m$^{2}$. Cherenkov light from nearby extensive air showers is 
focused onto the cameras that consist of 499 photomultiplier tube (PMT) pixels. 
The spacing between the PMT pixels corresponds to 0.15$^{\circ}$ on the sky, resulting in a total camera field of view of 3.5$^{\circ}$. 
VERITAS has a large 
effective area (greater than 10$^{5}$ m$^{2}$ above 1 TeV), an energy resolution of 15-20\% and a 
single-event angular resolution of 0.08$^{\circ}$ (for energies greater than 500 GeV). 
This enables the detection of a point source flux 
that is 1\% of the Crab nebula at a 5$\sigma$ significance in less than 30 hours 
\citep{2009arXiv0912.5355O}. For more details on the VERITAS instrument, see \citet{2008AIPC.1085..657H}.

VERITAS employs a three-level trigger system~\citep{2008ICRC....3.1539W}. At the pixel level there 
must be a signal greater than 50 mV 
(3-4 photoelectrons produced by $\sim$ 20 photons), which is monitored by a constant fraction 
discriminator. An individual camera triggers when at least three adjacent pixels meet the first condition 
within a 5 ns time window. Finally, two or more telescopes must trigger within 50 ns for the event to 
be recorded. The photomultiplier tube relative gains are calculated using dedicated LED flasher 
calibration runs taken nightly \citep{2010NIMPA.612..278H}. Recorded showers are then parametrized by their 
second order moments, commonly referred to as Hillas parameters \citep{1985ICRC....3..445H}.

The data presented here were collected between 2009 and 2012. In order to reduce the systematic 
uncertainties in the background determination, observations were made using the wobble technique 
~\citep{1994APh.....2..137F,2007A&A...466.1219B}.  In 2009, all data were collected with the telescopes pointing to regions in 
the sky offset 0.5$^\circ$ from the HEGRA source position. In later seasons almost all the data from 2010 
until 2012 were collected centering the telescopes on points equidistant from both \tev\ and Cygnus 
X-3 \footnote{Cygnus X-3 is 0.5$^\circ$ south of \tev. } (see Figure \ref{fig:sigmap}).  A 
small amount of data was taken with the telescopes aimed at four wobble positions offset 0.5$^\circ$ from 
Cygnus X-3.

A total of 48.2 hours of data was selected after removing data taken under bad 
weather conditions. \tev\ was observed with a mean telescope elevation of 68$^{\circ}$. The data were 
analyzed using the standard VERITAS calibration and reconstruction tools \citep{2008ICRC....3.1325D}. 
Images from all participating telescopes in an event are combined to obtain the parameters of the arriving 
gamma-ray \citep{1997JPhG...23.1013F,2006APh....25..380K}. 
In order to suppress the large number of cosmic ray background events, gamma-ray/hadron separation criteria 
(cuts) are employed that compare the shapes of the shower images with those from simulated gamma-ray images. 
The results presented here required at least three telescopes to have recorded images of the shower with 
more than 1000 photons in each image. 
An additional cut on the square of the arrival angle of the incoming gamma ray with respect 
to the source position ($\theta^{2}$ $<$ 0.055) is applied to extract the signal. The ring background 
model has been used to estimate the background \citep{2007A&A...466.1219B}. 
The cuts have been previously optimized using a simulated source that has a flux 5\% of the 
flux of the Crab nebula, and whose spectrum is at least as hard as the Crab nebula. 
The resulting analysis threshold is 520 GeV, which corresponds to the peak of the reconstructed energy 
distribution. 
A significance is calculated 
using the surviving gamma-ray like events and equation 17 in \citet{1983ApJ...272..317L}.

\section{VERITAS Results} 
\label{sec:results}

\begin{figure}
  \begin{center}
     \includegraphics[width=25.0pc]{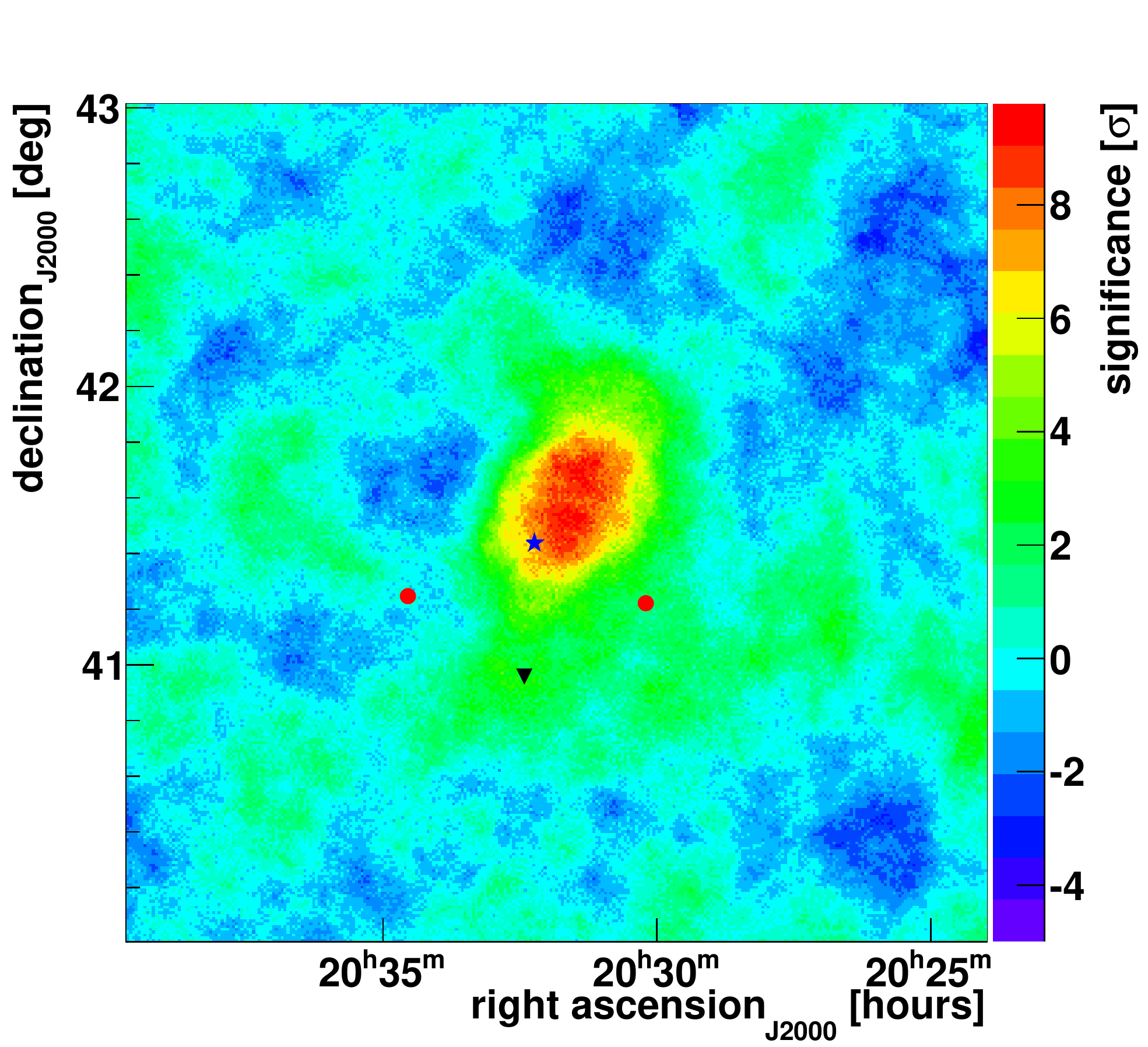}
  \end{center}
  \caption{VERITAS gamma-ray significance map centered on \vtev. The color indicates the significance within an integration
window of 0.23$^{\circ}$. The blue star marks the location of the $Fermi$-LAT pulsar PSR J2032+4127, the black triangle 
is the location of Cygnus X-3 and the red circles represent the shared wobble locations.}
\label{fig:sigmap}
\end{figure}

The analysis resulted in 595 events recorded in the source region and 3054 events in the 
selected background region. The geometrical background region selected was 7.7 times larger 
than the source region. This yields a significance of 8.7 standard deviation standard deviationss at the HEGRA position of \tev. 
Figure \ref{fig:sigmap} shows the VERITAS significance sky map. The gamma-ray point spread 
function for the analysis is 0.08$^{\circ}$. The map is smoothed with a top-hat function corresponding to 
the source integration radius of 0.23$^{\circ}$. The morphology of the source is investigated by 
binning the uncorrelated acceptance-corrected map of excess events.  The uncorrelated excess map is 
fit to a two-dimensional Gaussian distribution in order to estimate the extent of the emission. 
The $\chi^{2}$ for the fit is 572.2 for 435 degrees of freedom.
It is found to be asymmetric with an extension along the major axis of 9.5$^{\prime}\pm$1.2$^{\prime}$, oriented 
to the North-West by 63$^{\circ}$$\pm$6$^{\circ}$, and 4.0$^{\prime}\pm$0.5$^{\prime}$ along the minor axis, 
the quoted errors are statistical. The 
position of the centroid was found to be 20$^{h}$ 31$^{m}$ 40$^{s}$ $\pm$ 65$^{s}$ and 41$^{\circ}$ 
33$^{\prime}$ 53$^{\prime \prime}$ $\pm$ 37$^{\prime \prime}$, which is consistent within errors with previous 
measurements, and we assign the name \vtev. 

\begin{figure}
  \begin{center}
     \includegraphics[width=25.0pc]{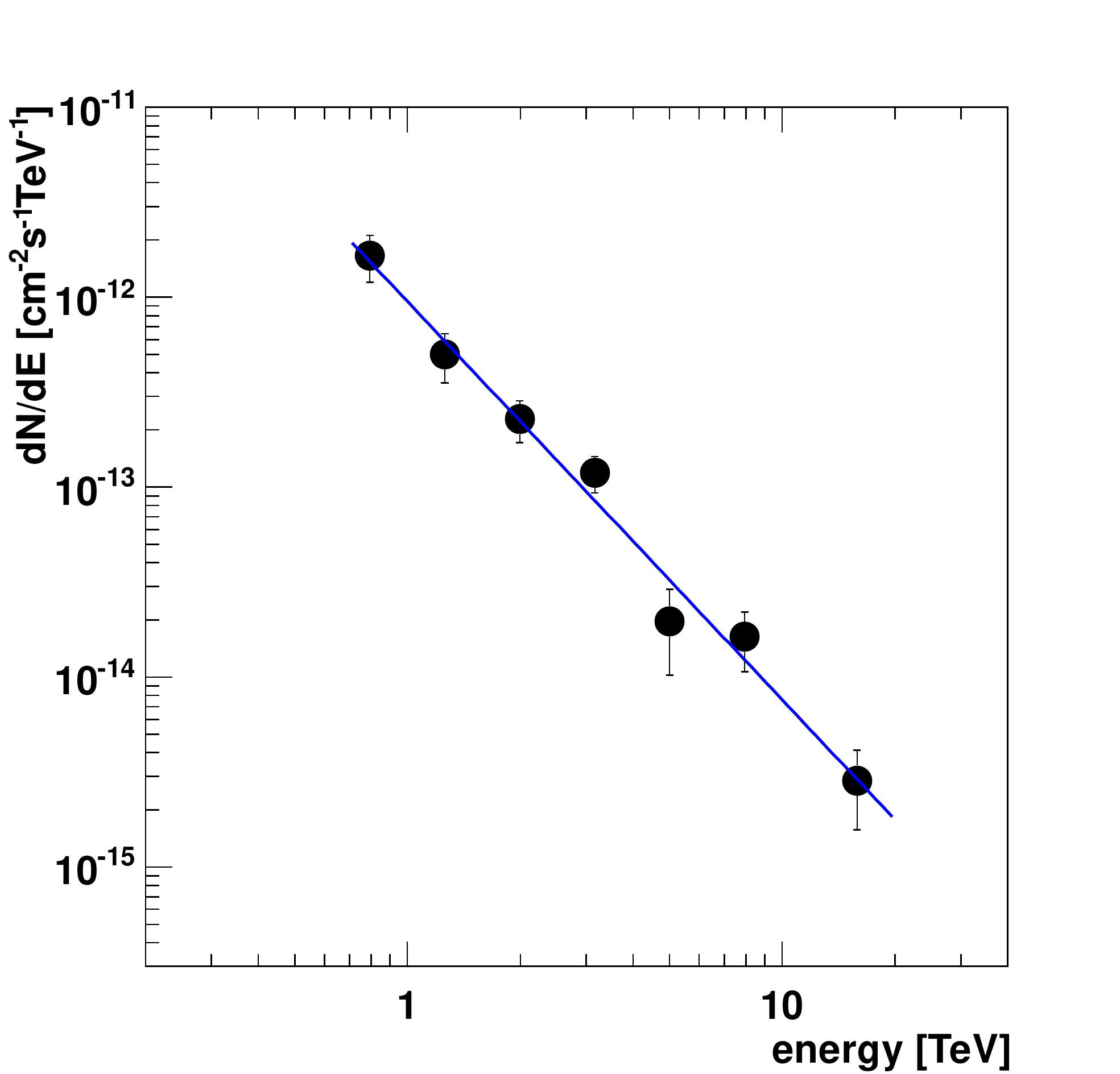}
  \end{center}
  \caption{VERITAS differential gamma-ray spectrum of \vtev. The blue line shows the power-law fit to the data points. A list of flux points
can be found in Table \ref{tab:spectra}.}
\label{fig:spectra}
\end{figure}

The spectrum, shown in Figure \ref{fig:spectra}, is well fit by a power law with a spectral index of 2.10 
$\pm$ 0.14$_{stat}$ $\pm$ 0.21$_{sys}$ and normalization at 1 TeV of (9.5 $\pm$ 1.6$_{stat}$ $\pm$ 
2.2$_{sys}$) $\times$ 10$^{-13}$ \spec. The $\chi^{2}$ per degree of freedom is 4.5/5 = 0.9.
We found no evidence of a cut-off up to 20 TeV. The total 
integrated flux above 1 TeV is 4.3\% of the Crab nebula flux. This corresponds to a flux of (2.35 $\pm$ 
0.55) $\times$10$^{-12}$ erg cm$^{-2}$ s$^{-1}$ above 1 TeV and represents 0.3\% of the spin-down 
luminosity of the pulsar PSR J2032+4127 (with an $\dot{E}$ of 2.7$\times$10$^{35}$ erg s$^{-1}$). Results 
of the spectral calculation can be seen in Table \ref{tab:spectra}.

\begin{figure}
 \begin{center}
  \includegraphics[width=13.5pc]{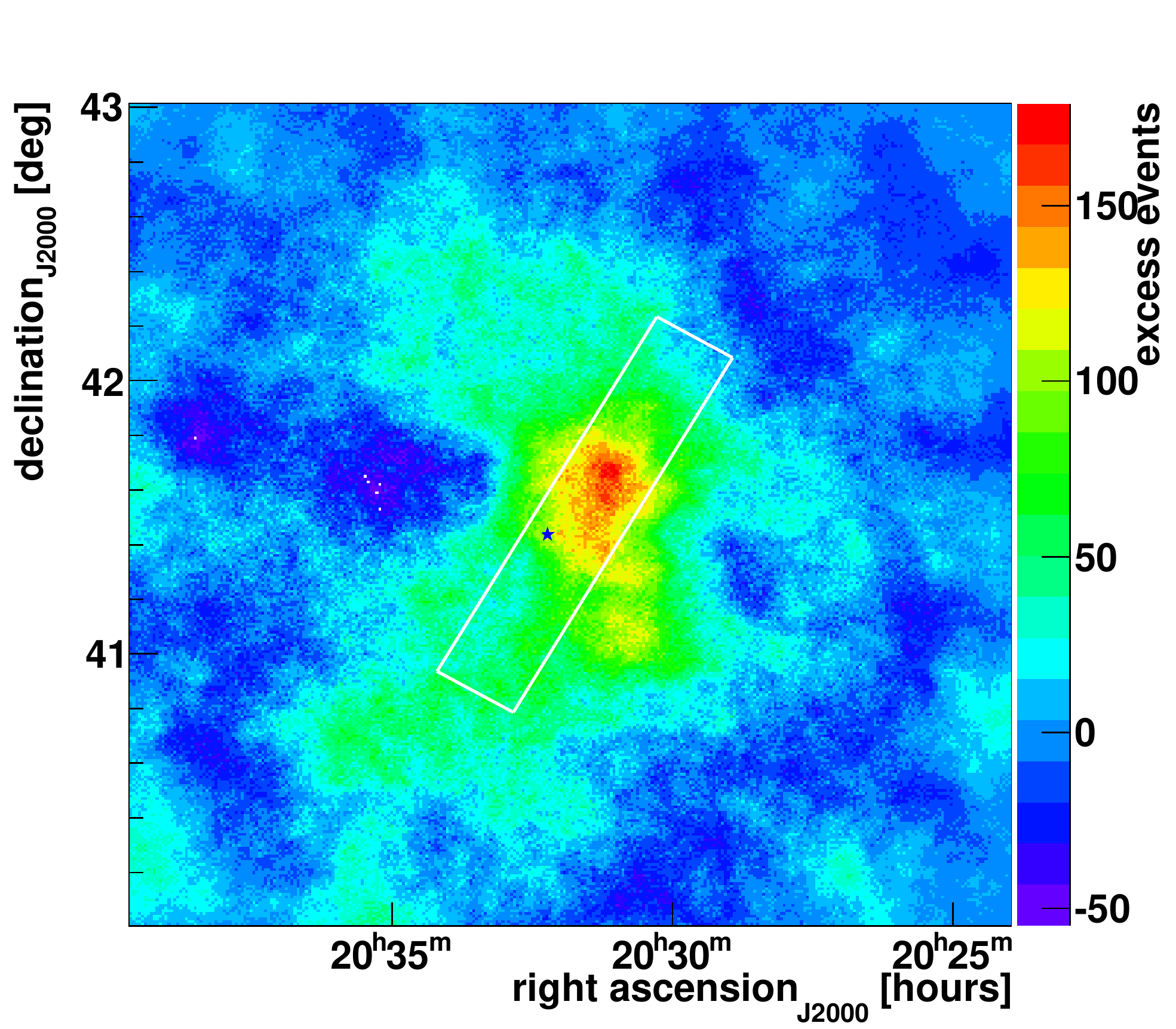}
  \includegraphics[width=13.0pc]{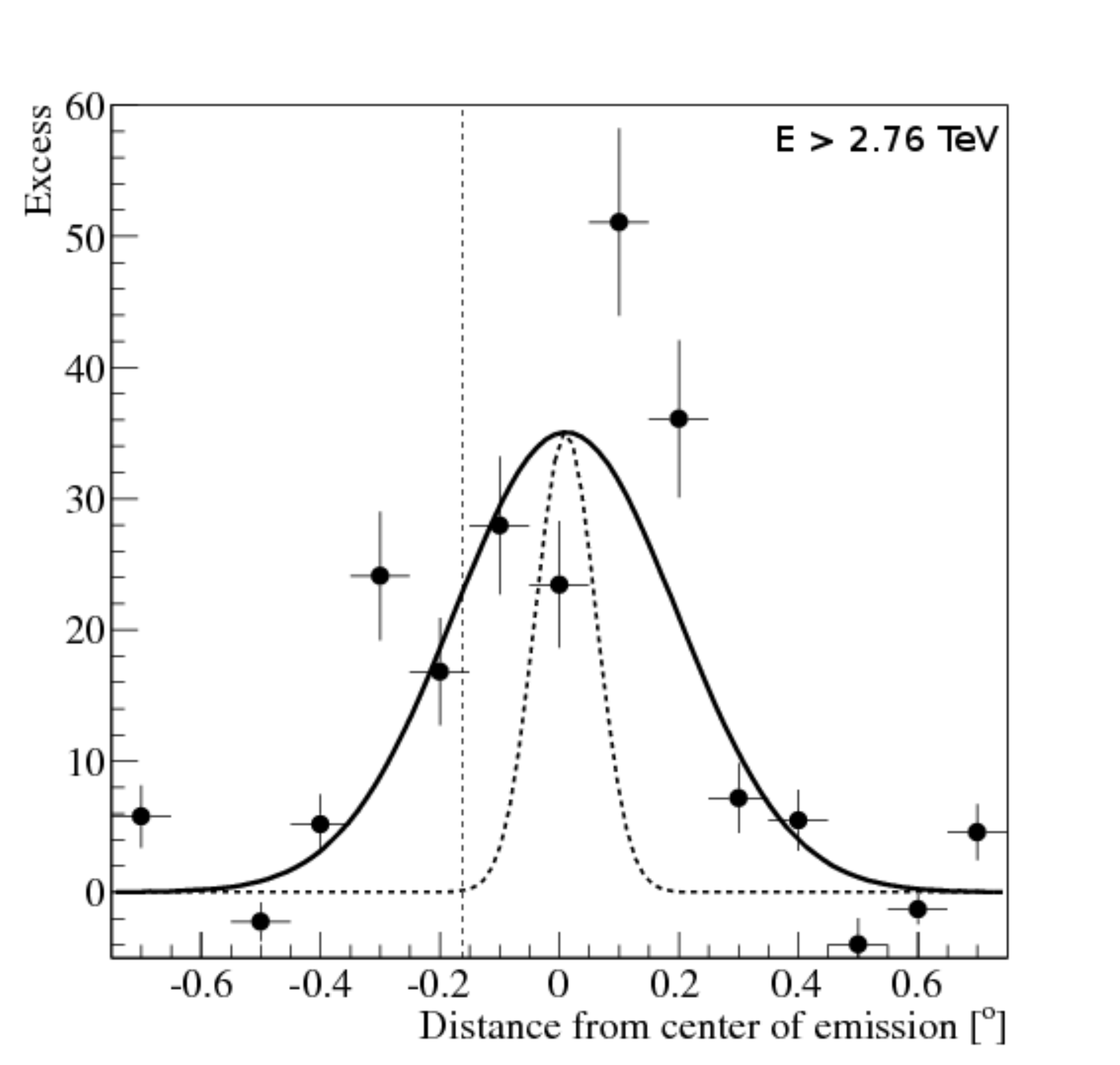}
  \includegraphics[width=13.5pc]{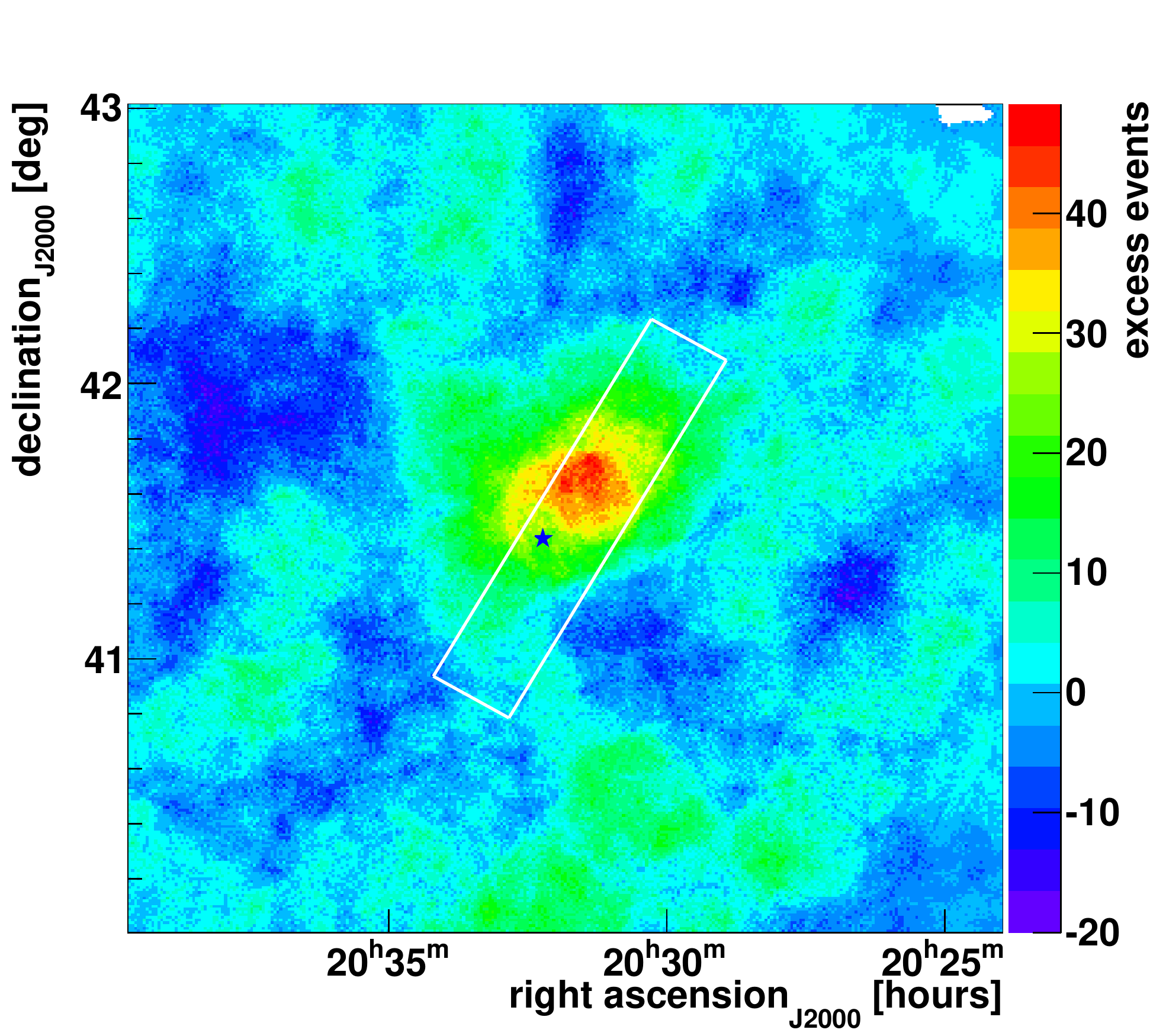}
  \includegraphics[width=13.0pc]{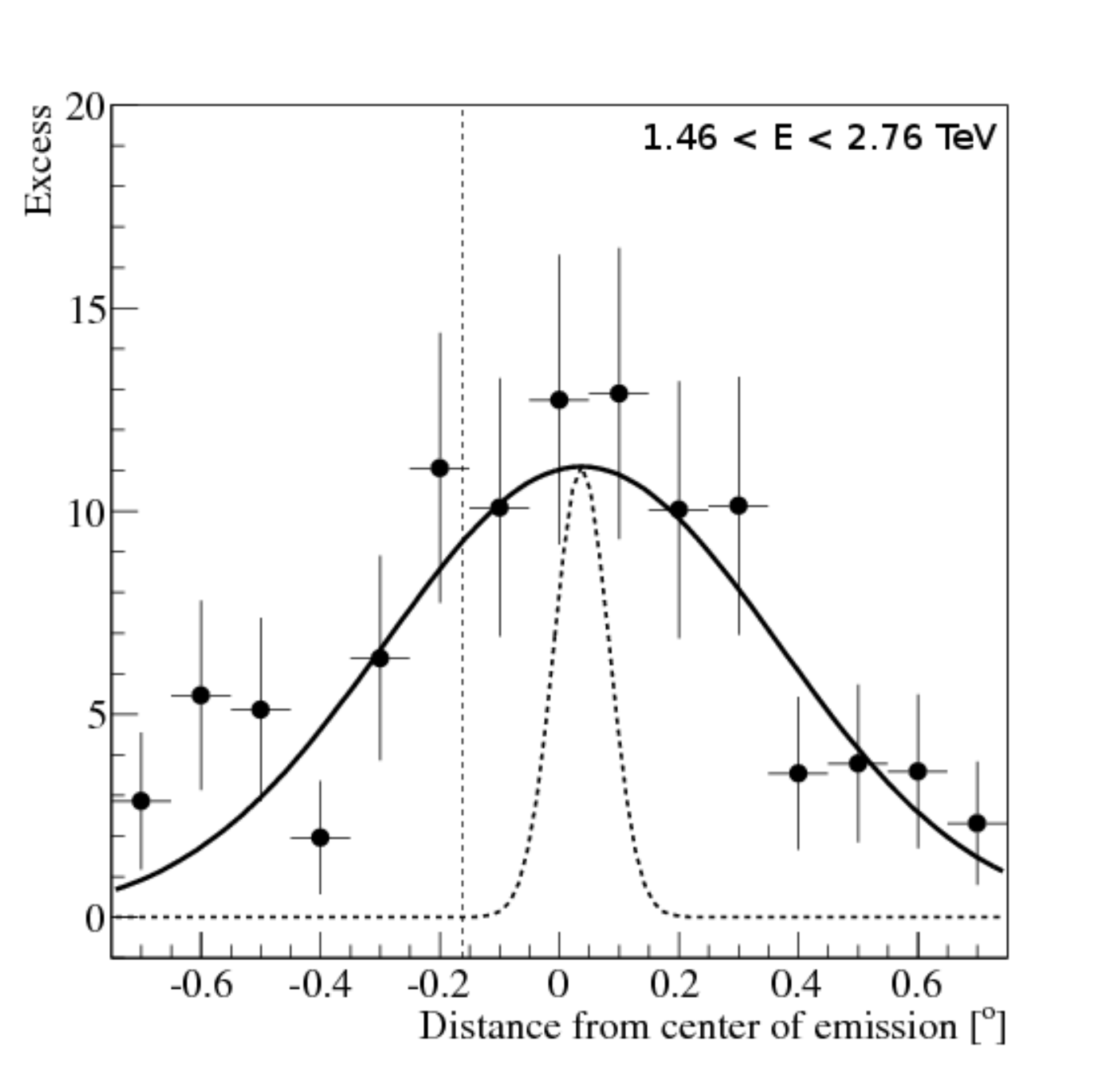}
  \includegraphics[width=13.5pc]{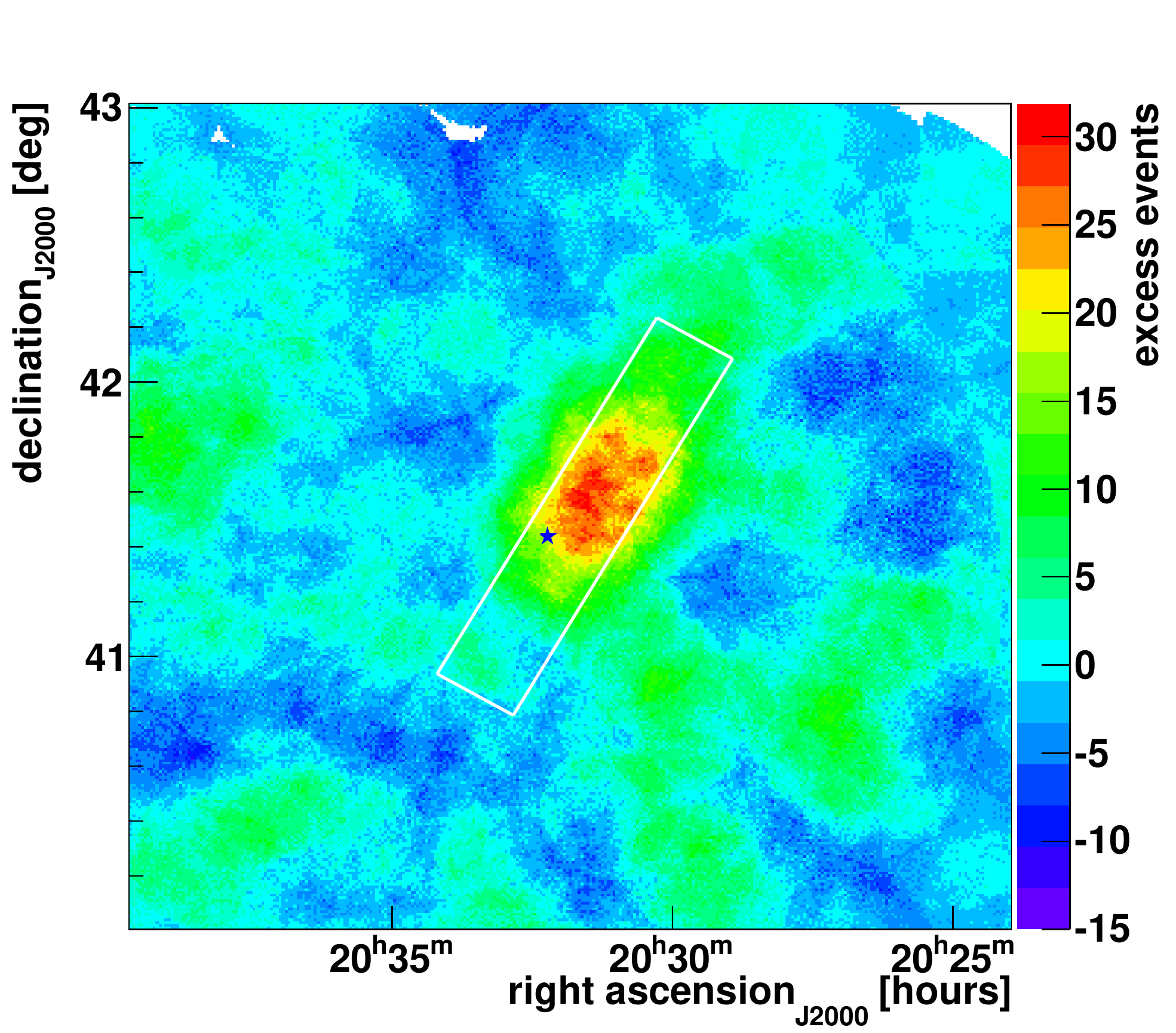}
  \includegraphics[width=13.0pc]{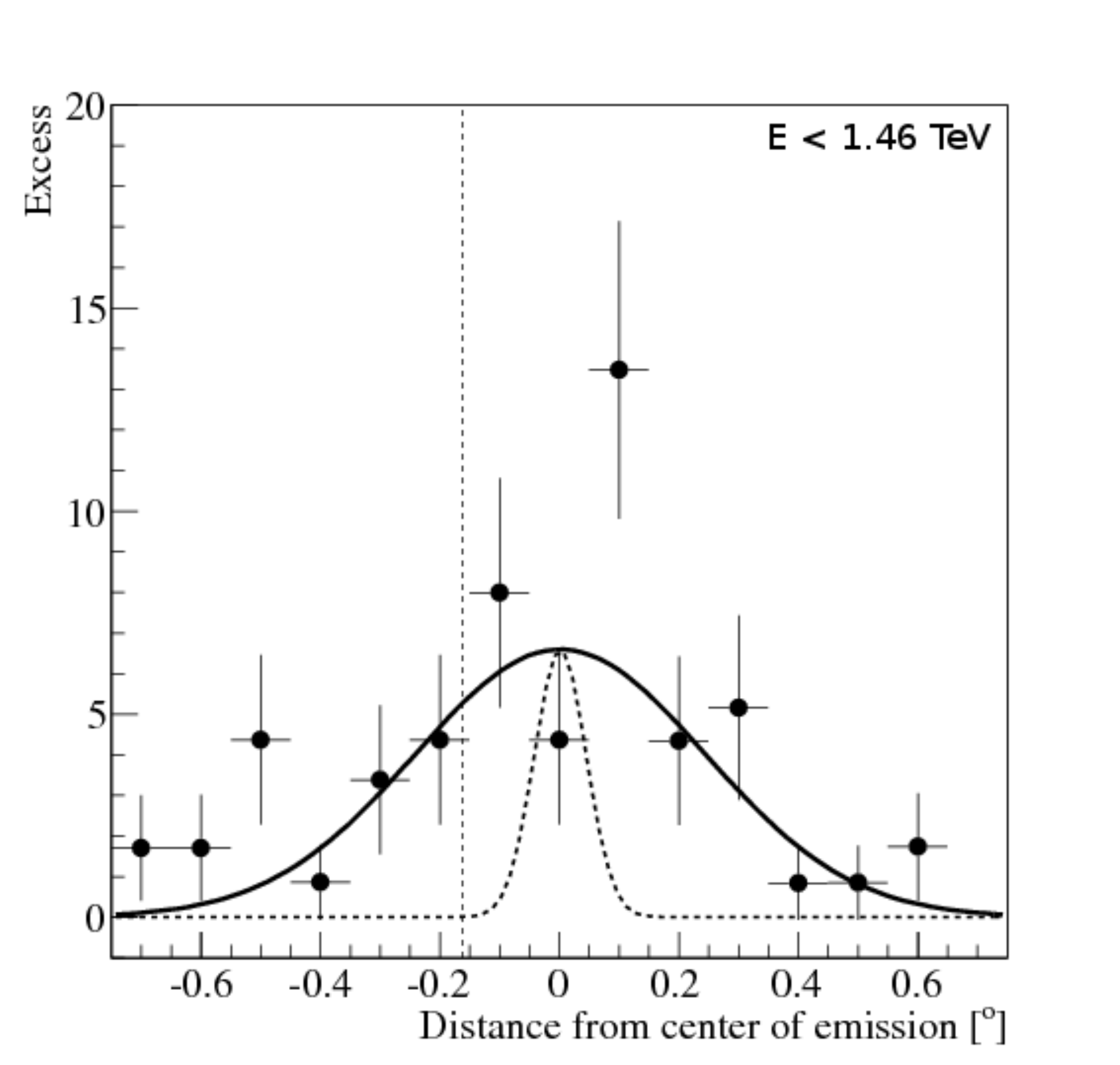}
 \end{center}
  \caption{
\small
\textit{Left}: correlated excess maps in the energy bands less than 1.46 TeV, 1.46 to 2.72 TeV and above 2.72 TeV. The white boxes
represents a cross-section aligned with the major axis of the excess. The $Fermi$-LAT pulsar PSR J2032+4127 is marked by the blue star.
\textit{Right}: slices of the uncorrelated excess map. The excess is fitted to a Gaussian distribution shown by the solid black line. 
Also plotted is the instrument response to
a point source following the same analysis procedure (dotted line). The \textit{vertical dotted line} represents the 
position of the $Fermi$-LAT pulsar. See text for details.
\large
}
 \label{fig:xsection}
\end{figure}

Figure \ref{fig:xsection} shows the VERITAS significance sky maps for three different energy ranges: E $<$ 
1.46 TeV, 1.46 $<$ E $<$ 2.7 TeV and E $>$ 2.7 TeV. The energy division was chosen a priori 
in an attempt to have equal statistics in each bin. Also shown is the one-dimensional 
histogram of the uncorrelated excess events along the major axis of the emission, as shown by the white boxes (which 
are 1.5$^{\circ}$ $\times$ 0.3$^{\circ}$ in extent). A Gaussian is fit to the excess. 
The dotted lined Gaussian is a set of Crab nebula data analyzed using the same analysis chain. 
This represents the response of the analysis to a point source. From this information the intrinsic width 
can be unfolded.
The Crab Nebula is a strong source compared to \tev ; 
a check has been made by reducing the Crab Nebula excess events by a factor of 20 and no significant change 
in the point source response was observed.
The results can be seen in Table \ref{tab:xsection}; based on the derived intrinsic width
no energy dependent morphology is observed.

\begin{small}
\begin{table}
\begin{center}
\begin{tabular}{c|c|c|c}
  \hline
    Energy  & Bin Width & Flux     & Significance  \\
    (TeV)   & (TeV)     & (\spec)  & ($\sigma$) \\
  \hline
    0.79  & 0.37 & (1.7 $\pm$ 0.5)$\times$10$^{-12}$ & 4.0 \\
    1.26  & 0.58 & (5.0 $\pm$ 1.5)$\times$10$^{-13}$ & 3.9 \\
    2.00  & 0.93 & (2.3 $\pm$ 0.6)$\times$10$^{-13}$ & 4.8 \\
    3.16  & 1.47 & (1.2 $\pm$ 0.3)$\times$10$^{-13}$ & 6.1 \\
    5.01  & 2.33 & (2.0 $\pm$ 0.9)$\times$10$^{-14}$ & 2.5 \\
    7.94  & 3.69 & (1.6 $\pm$ 0.6)$\times$10$^{-14}$ & 4.0 \\
    15.85 & 15.12& (2.9 $\pm$ 1.3)$\times$10$^{-15}$ & 2.8 \\
  \hline
\end{tabular}
\end{center}
\caption{Differential flux of \vtev ~measured by VERITAS. 
Also see Section \ref{sec:results} and Figure \ref{fig:spectra}.
The errors quoted are the 1 $\sigma$ statistical errors.
}
\label{tab:spectra}
\end{table}
\end{small}

\begin{small}
\begin{table}
\begin{center}
\begin{tabular}{c|c|c|c|c}
  \hline
    Energy  & Point Source & Total & Intrinsic   & $\chi^{2}$/degrees of freedom \\
    Range   & Width & Width        & Width       & \\
    (TeV)   & (degrees) & (degrees)  & (degrees) & \\
     & Dotted Line & Solid Line & & \\
  \hline
    $\leq$ 1.46  & 0.051  & 0.19$\pm$0.01 & 0.18$\pm$0.03  & 68.1 / 12 \\
    1.46 -- 2.76 & 0.046  & 0.33$\pm$0.05 & 0.33$\pm$0.10  & 12.4 / 12 \\
    $\geq$ 2.76  & 0.044  & 0.24$\pm$0.04 & 0.24$\pm$0.09  & 15.0 / 11 \\
  \hline
\end{tabular}
\end{center}
\caption{Gaussian fits to slices of the uncorrelated excess; see Figure \ref{fig:xsection}. 
Widths are defined as 1 standard deviation. 
The $\chi^{2}$ per degree of freedom for the fit to the total width is shown in the final column.
Note the error on the point source width is dominated by systematic errors. 
}
\label{tab:xsection}
\end{table}
\end{small}

\section{$Fermi$-LAT Analysis}

An analysis of the $Fermi$-LAT data from the region around \vtev\ has been performed in the energy range
500 MeV to 100 GeV. The $Fermi$-LAT is an electron-positron pair-conversion 
telescope \citep{2009ApJ...697.1071A}. The analysis described here uses data taken during the first four 
years of the operation of the $Fermi$-LAT detector. Data reduction was performed using the publicly 
available LAT data tools (version v9r27) and follows the analysis scheme described in 
\citet{2009ApJ...697.1071A}.  Initially, all events within a 20$^{\circ}$ region of interest around \vtev\ 
were selected for further analysis. The resulting data set were analyzed using a binned likelihood 
technique \citep{1979ApJ...228..939C, 1996ApJ...461..396M}. The routine is implemented in the LAT data 
tools as $gtlike$, which calculates the likelihood function probability using a source model folded with 
the LAT instrument response function (IRF, $P7 V6$). Finally a model of the source region was compared to 
the counts map to search for a signal.

The study of the associated diffuse source required us to take into account the pulsed emission of PSR 
J2032+4127 and assign phases to the gamma-ray photons and select only those in an off-pulse window, 
thereby minimizing contributions from the pulsar. Accurate timing solutions based on radio data from the 
Green Bank Telescope were used in conjunction with TEMPO2 \citep{2006MNRAS.369..655H} in order to phase 
fold the photon data.

For the timing analysis, photons between 100 MeV and 300 GeV and within 0.5$^\circ$ of the pulsar position 
R.A. = 20$^h$32$^m$13.1$^s$, Decl. = +41$^\circ$27$^\prime$ 24.6$^{\prime \prime}$ were selected. 
Following the pulsar analysis, the off-pulse region was identified (approximately 20\% of the total time 
was removed) and a binned likelihood analysis was again performed. No significant emission from the region 
of \vtev\ was observed and therefore we compute the 99\% upper limits \citep{2005NIMPA.551..493R}.
The assumed spectral index was 2.0 and the calculated results are shown in Table \ref{tab:fermiul}.

\begin{small}
\begin{table}
\begin{center}
\begin{tabular}{c|c|c}
  \hline
    Energy  Range & 99\% Upper Limit & 99\% Upper Limit \\
    (GeV)   & (photons cm$^{-2}$ s$^{-1}$) & (erg cm$^{-2}$ s$^{-1}$) \\
  \hline
    0.5 -- 1  & 5.73 $\times$ 10$^{-9}$  & 1.18 $\times$ 10$^{-11}$ \\
    1 -- 10   & 1.94 $\times$ 10$^{-9}$  & 1.12 $\times$ 10$^{-12}$ \\
    10 -- 100 & 2.06 $\times$ 10$^{-11}$ & 1.20 $\times$ 10$^{-13}$ \\
  \hline
\end{tabular}
\end{center}
\caption{Upper limits on gamma-ray emission from the region of \tev\ from the analysis of $Fermi$-LAT data.}
\label{tab:fermiul}
\end{table}
\end{small}

\section{Multiwavelength Properties and Interpretations}

Ever since its discovery by HEGRA \citep{2002A&A...393L..37A}, multiple observational efforts have been 
undertaken to identify the potential counterparts of \tev\ at other wavelengths. However, after years of 
multiwavelength observations, the origin of gamma-ray emission from the region still remains unresolved.

\begin{figure}
  \begin{center}
     \includegraphics[width=35.0pc]{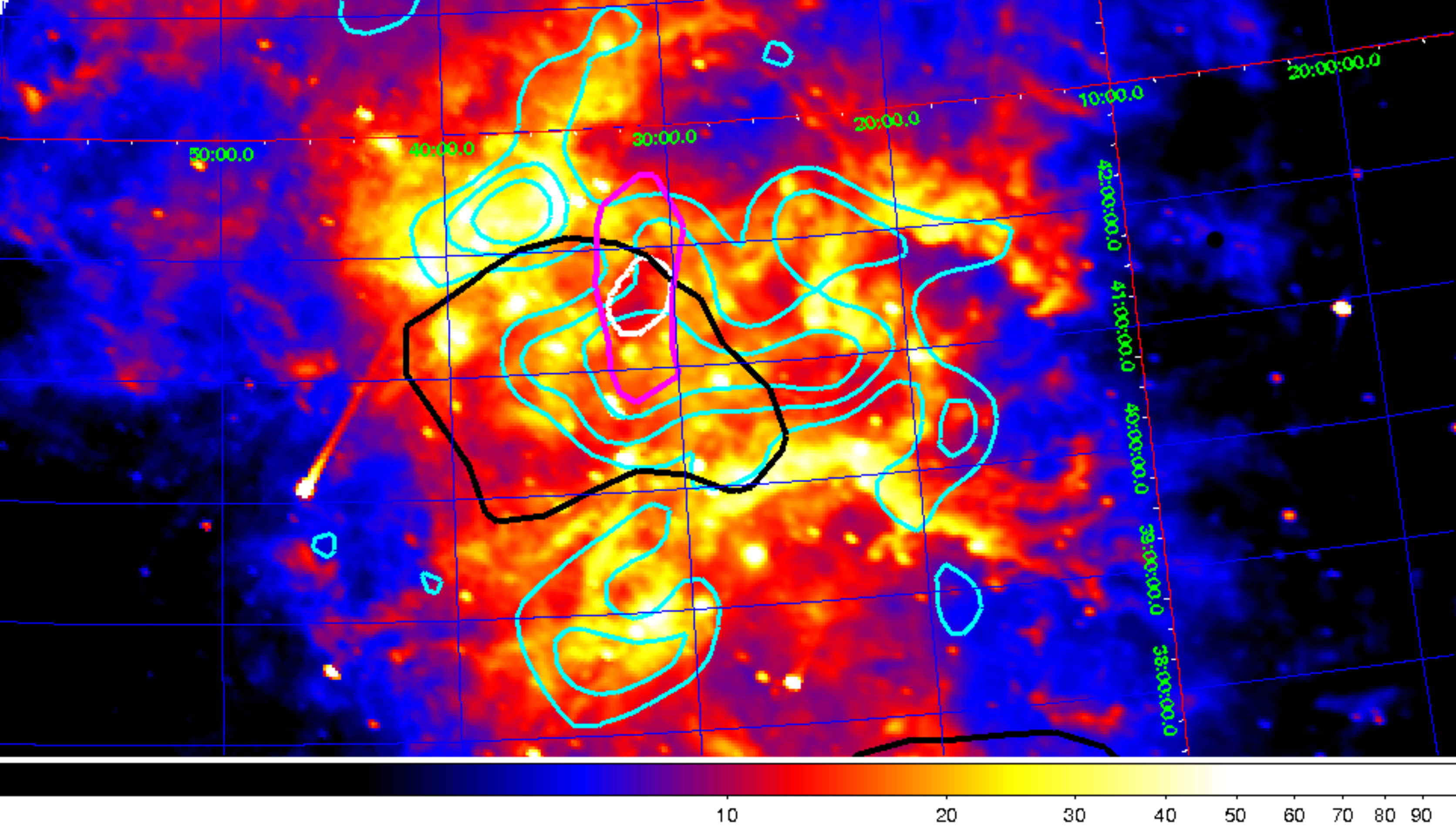}
  \end{center}
  \caption{MSX 8 $\mu$m infrared survey (color, W m$^{-2}$ sr$^{-1}$, in log scale) for the region around \vtev. 
The black, magenta and white contours represent the Milagro, ARGO and VERITAS 5 standard deviation level.
The $Fermi$-LAT 0.16, 0.24 and 0.32 photons/bin contours are shown in cyan.
}
 \label{fig:mgro}
\end{figure}

The Milagro water Cherenkov detector has performed a large-scale survey of the Cygnus region and has 
discovered a population of extended sources without compelling counterparts \citep{2007ApJ...658L..33A}. 
Figure \ref{fig:mgro} shows the 8 $\mu$m Midcourse Space Experiment (MSX) map of the region with the 
Milagro and VERITAS 5$\sigma$ contours overlaid in black and white respectively. The VERITAS flux has been 
integrated over a circular region with radius of 0.23$^{\circ}$ whereas, the Milagro flux has been derived 
over a region of 3$^{\circ}$ $\times$ 3$^{\circ}$, owing to the larger angular resolution of the detector. 
It is of course quite possible that the flux of gamma rays measured by Milagro is not exclusively from \tev\ but 
also contains a significant diffuse (or yet unresolved sources in the region) component. 
These observations are joined by those of the ground array detector ARGO 
\citep{2012ApJ...745L..22B}.  
MAGIC and VERITAS measure spectral indices of 2.0 $\pm$ 0.3$_{stat}$ and 2.10 $\pm$ 0.14$_{stat}$ respectively.
In comparison both the Milagro and ARGO observations resulted in significantly softer spectral indices of
3.2$\pm$0.2 and 2.8$\pm$0.4.
ARGO operates at comparable energies to those of the imaging Cherenkov telescopes
and sees an extension that is twice as large.

At GeV gamma-ray energies $Fermi$-LAT has reported the
existence of a cocoon located in the Cygnus region, spread over 2$^{\circ}$ between \vtev\ and
$\gamma$-Cygni. Even though pulsar wind nebulae are known to power extended gamma-ray sources,
it is unlikely that the cocoon is powered by PSR~J2032+4127 due to its size.
\citet{2011Sci...334.1103A} suggest that the gamma-ray excess of the cocoon
is due to a population of freshly accelerated cosmic rays.
The link between the cocoon and the emission at TeV energies remains unclear and we cannot rule out any connection between the two.

\begin{figure}
 \begin{center}
  \includegraphics[width=17.pc]{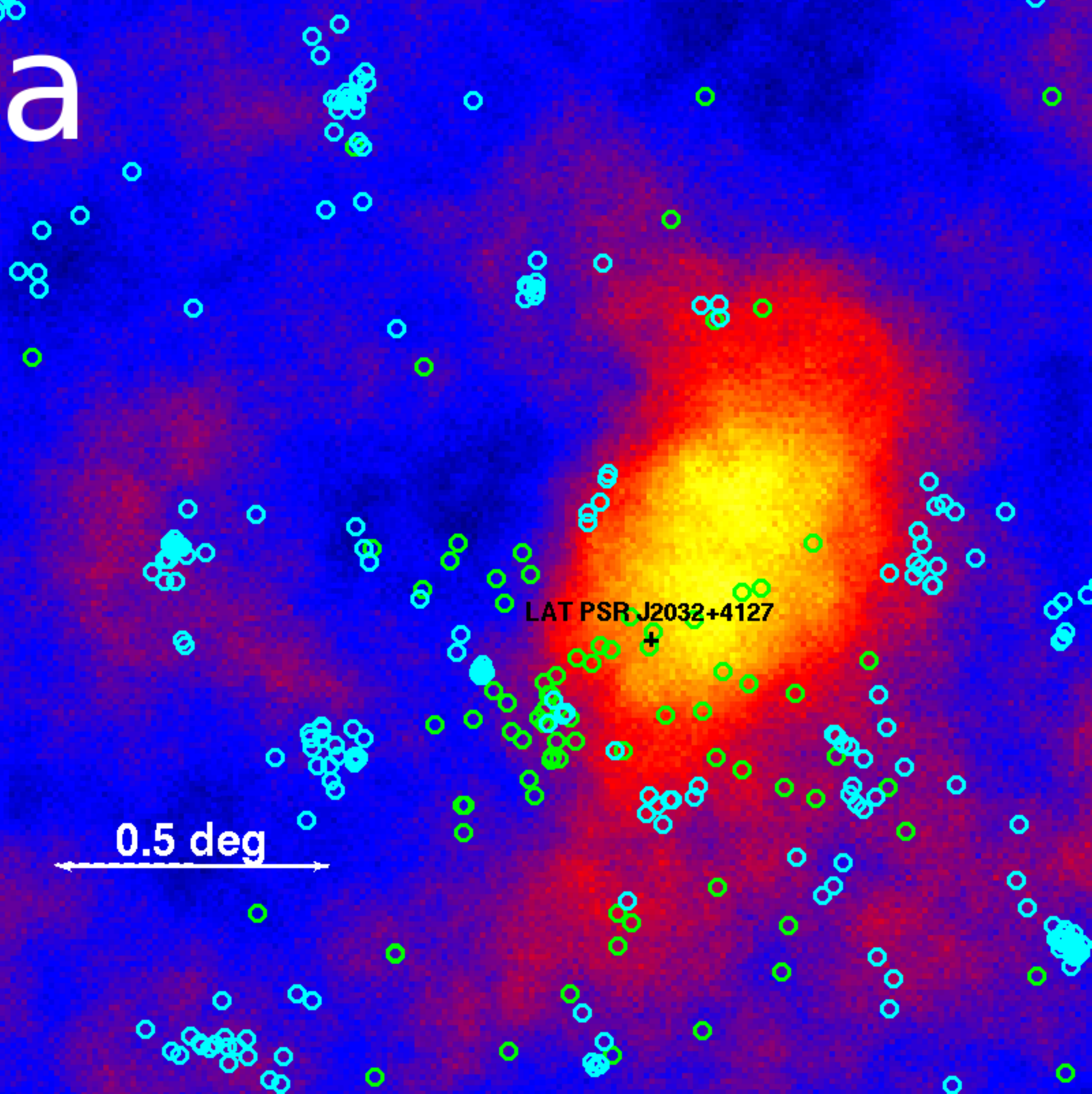}
  \includegraphics[width=17.pc]{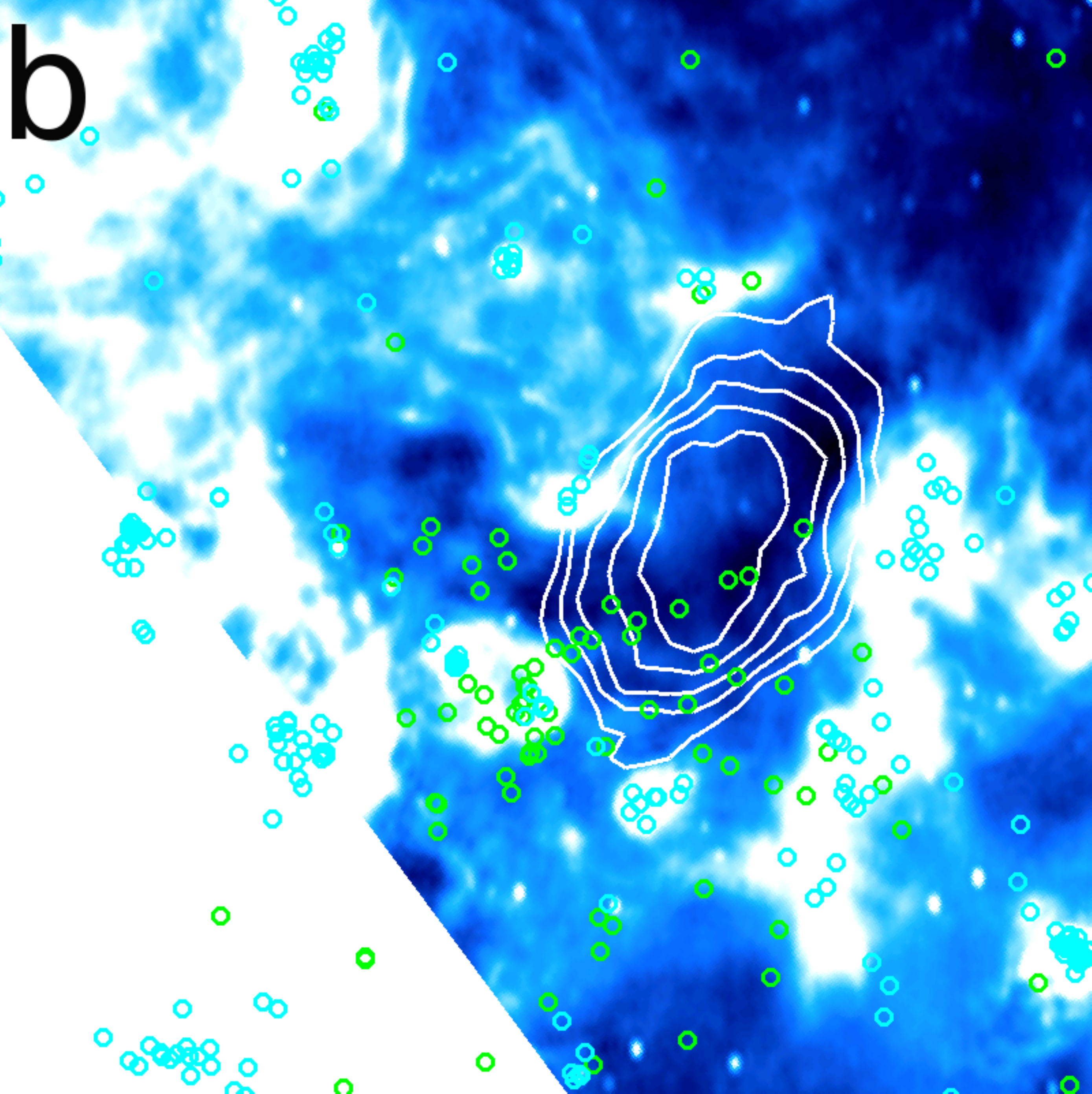}
  \includegraphics[width=17.pc]{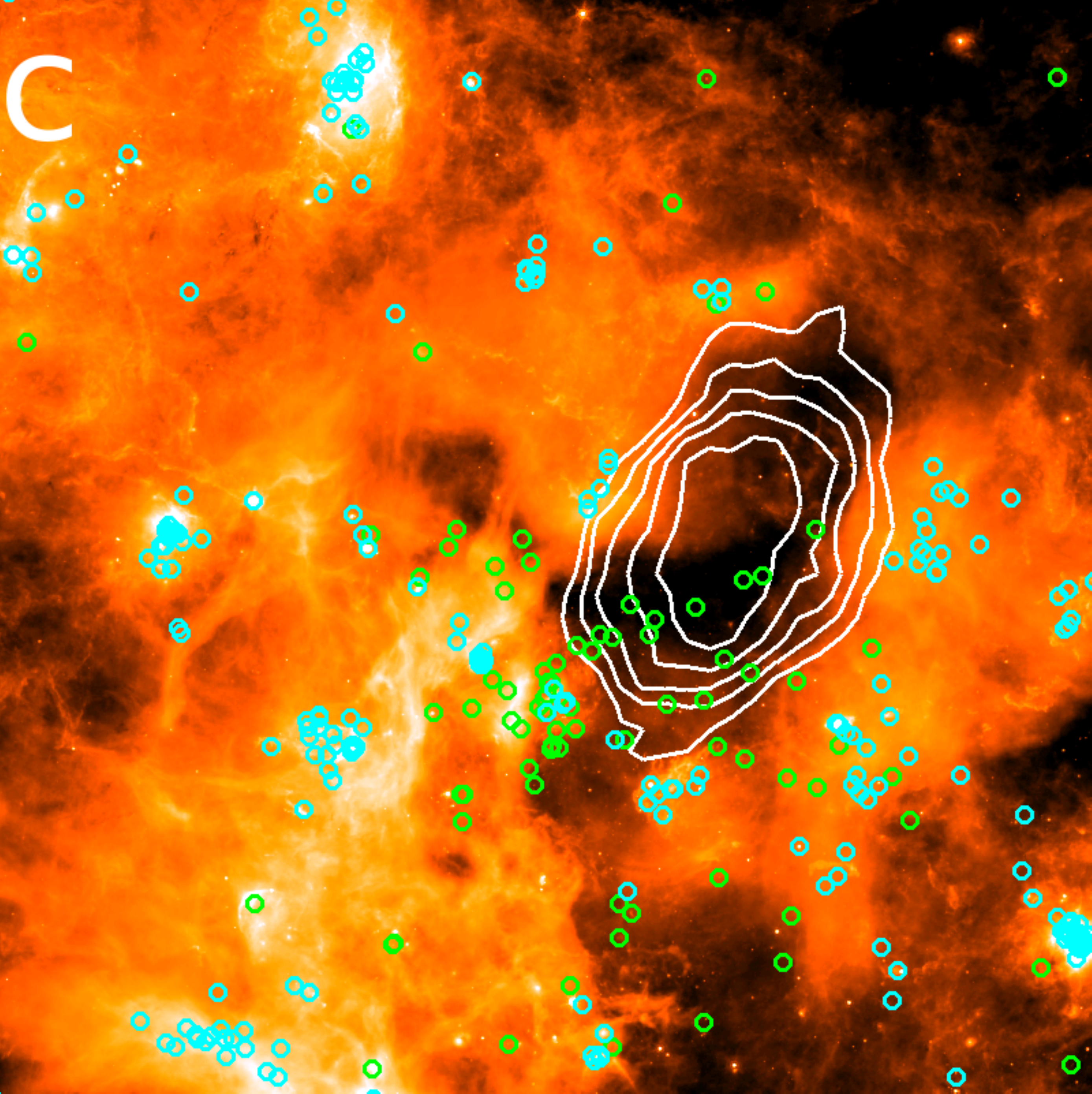}
  \includegraphics[width=17.pc]{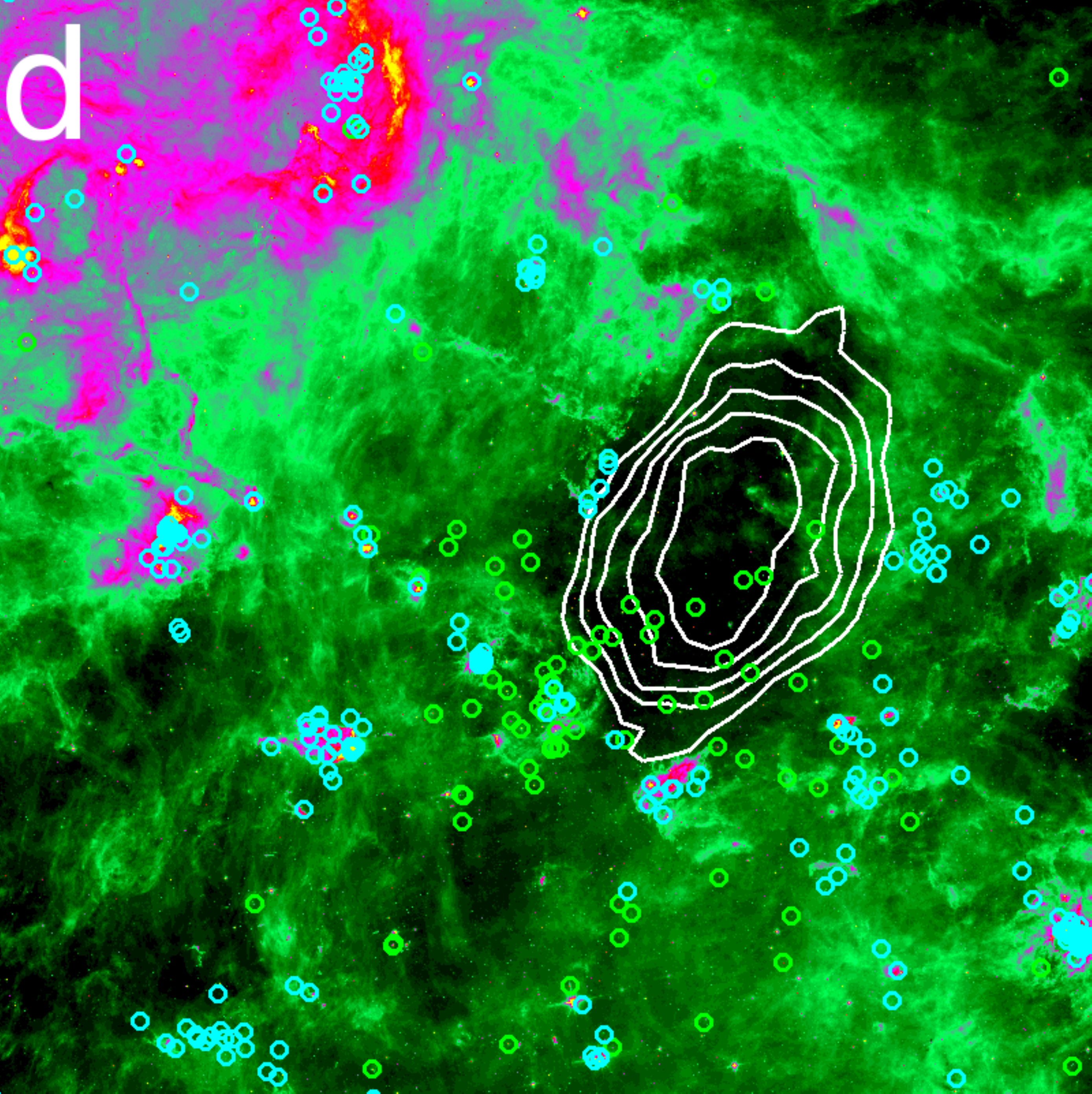}
 \end{center}
 \caption{
\vtev\ and its vicinity at different wavelengths. a: VERITAS significance map with the position of $Fermi$-LAT 
PSR J2032+4127 indicated by a black cross.
b: 1.4 GHz image from the Canadian Galactic Plane Survey (CGPS; \citet{2003AJ....125.3145T}). c: $Spitzer$ MIPS 
24 $\mu$m image from the MIPSGAL survey \citep{2004ApJS..154...25R}. d: $Spitzer$ GLIMPSE 8 $\mu$m image \citep{2009PASP..121..213C,2003PASP..115..953B}. In images
b, c, and d the VERITAS significance contours from 4 to 8 standard deviations are shown as white curves. 
Green circles are OB stars \citep{2012A&A...543A.101C}. Cyan circles are
star forming regions \citet{2002yCat.5112....0A}.
}
 \label{fig:mwplot}
\end{figure}

Figure \ref{fig:mwplot} shows multiwavelength images of \vtev\ and its vicinity. Assuming a distance of 
1.7~kpc, the one $\sigma$ width of the TeV gamma-ray emission is 4.7 $\times$ 2.0 pc along the major and 
minor axes. Due to the star-forming activity known to take place in Cygnus X, one of the richest known 
regions of star formation in the Galaxy, the infrared (IR) and radio images are dominated by bright 
diffuse emission exhibiting complex and intricate structure. Interestingly, nearly all the TeV gamma-ray 
emission happens to be confined within one of the rare $voids$. Although a chance coincidence is possible, 
the rarity of these voids and the similarity between the TeV source morphology and that of 
the void hint at a possible physical connection between the two.  The large size of the void and lack of a 
characteristic patch structure make it dissimilar to the dark infrared clouds (composed of cold molecular 
material) that were recently found within the Cygnus X complex \citep{2009AAS...21335601H}. 
Moreover, CO line imaging, see 
\citet{2003ApJ...597..494B}, shows that compact CO emission is only seen in the eastern part of 
the void and of the extended TeV source. Hence, it does not seem likely that the void is due to absorption 
by cold molecular material along our line-of-sight.

One possibility is that the void is formed due to the collective action of powerful stellar winds from 
an association of massive stars, a hypothesis considered by \citet{2003ApJ...597..494B} and 
\citet{2008ApJ...675L..25A}. They argued that the presence of a large, mechanical power density from the 
stellar winds of the OB stars, make Cygnus OB2 a prime candidate for the investigation of the stellar wind 
hypothesis for the acceleration of Galactic cosmic rays. Massive stars are common in Cygnus OB2 and the 
energy required to power the VHE emission is only a fraction of the estimated wind kinetic energy. 
However, many of these massive stars are outside the observed TeV gamma-ray emission region (see cyan 
circles in Figure \ref{fig:mwplot}, top panels). Thus the observed TeV gamma-ray morphology seems 
unlikely to be produced by massive stellar winds.

\begin{figure}
 \begin{center}
  \includegraphics[width=5in,angle=0]{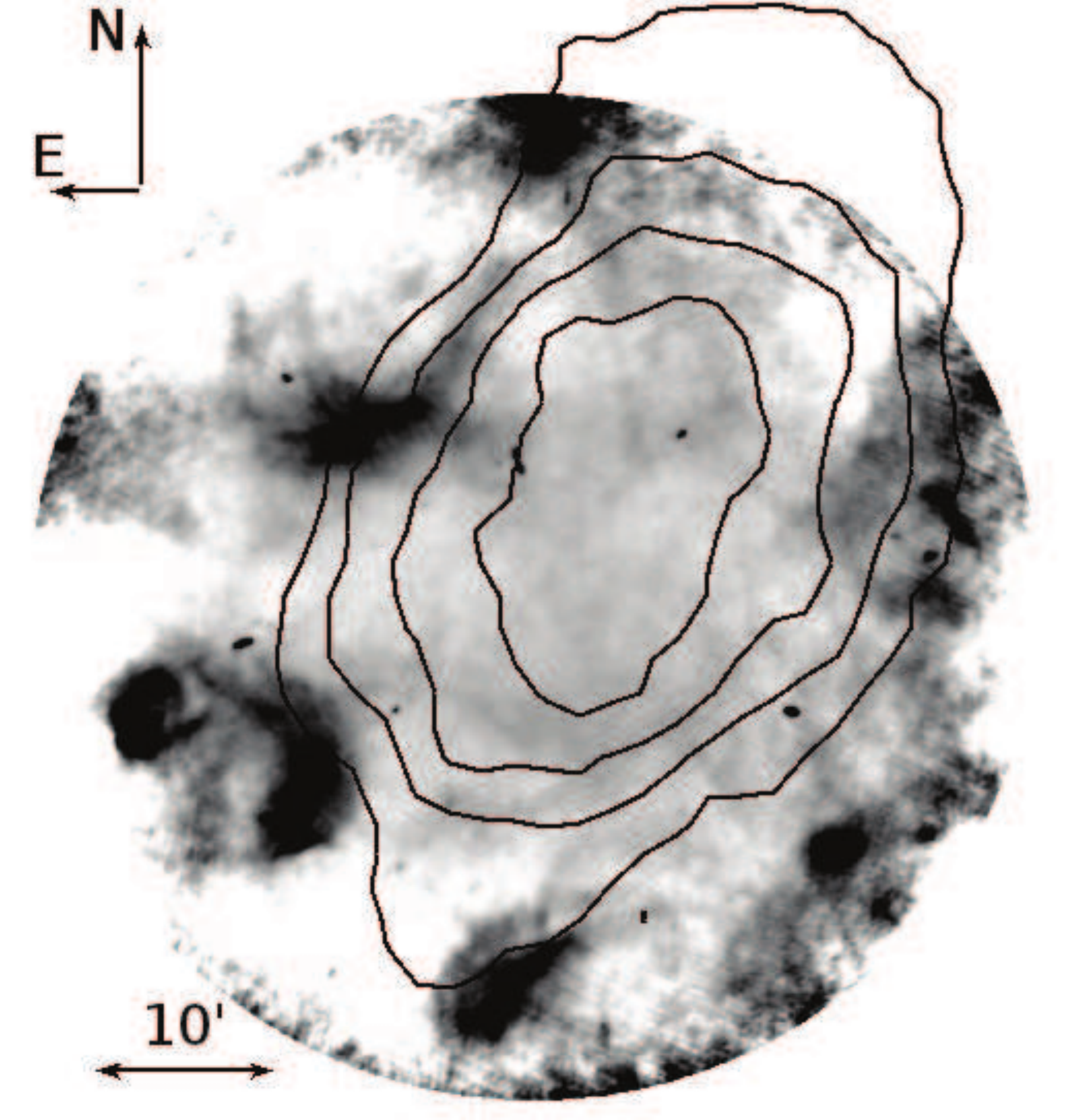}
 \end{center}
 \caption{ \footnotesize{VLA 20-cm (L-band) image with VERITAS 4 to 8 $\sigma$ contours overlaid \citep{1998AJ....115.1693C}.}}
 \label{fig:vla}
\end{figure}

It is possible that the void is due to a supernova explosion in Cygnus OB2 resulting in a SNR that 
expanded into the surrounding medium. \citet{2008MNRAS.385.1764B} mention hints of a shell-like 
structure with a radius of $\sim5'$ seen in the 6 cm VLA image.  However, we note that this size is 
smaller than the size of the void and the TeV gamma-ray source extent. Although some faint non-thermal 
radio emission is present within \vtev\ (see Figure \ref{fig:vla}), it may be due to a PWN within the SNR 
rather than the SNR shell.

If the SN explosion occurred relatively long ago ($\gtrsim$ 30,000 years), the shell could have grown much 
larger and become fainter. In this scenario the TeV gamma-ray emission would come from the interior of 
the aged SNR filled with a pulsar wind. It would then be natural to assume that the SNR is linked to LAT 
PSR~J2032$+$4127, which is apparently accompanied by a faint X-ray PWN. The faintness of the X-ray PWN 
could be attributed to the large spin-down age of the pulsar and comparatively low $\dot{E}$. Based on a 
distance of 1.7 kpc the gamma-ray efficiency is $\epsilon_{\gamma}$ = $L_{\gamma}$/$\dot{E}$ = 0.3\% (in 
the energy range 1 - 10 TeV), while the X-ray PWN efficiency is only 0.04\% (between 0.5-8.0 keV). 

The position of the pulsar suggests that it might be moving South-East along the elongation of the TeV 
gamma-ray emission. Given the observed angular separation between the center of emission and the pulsar, 
the transverse velocity is calculated to be 51 km s$^{-1}$, assuming a characteristic age of 0.11 Myr and 
a distance of 1.7 kpc. 

\begin{figure}
\begin{center}
\includegraphics[width=37.0pc]{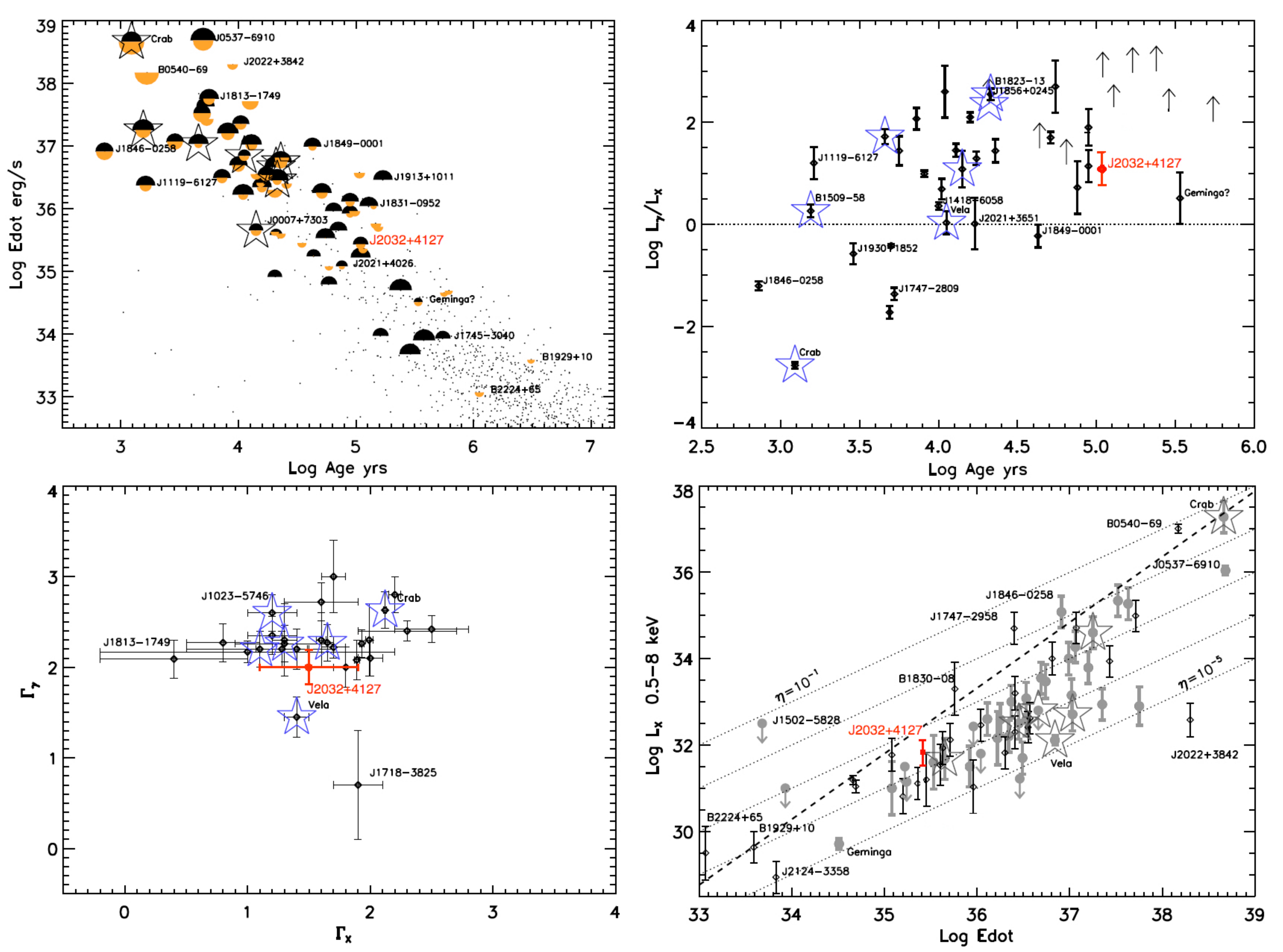}
\end{center}
\caption{
{\em Top left:} Pulsars with detected PWNe (or PWN candidates) in the $\tau_{\rm sd}$-$\dot{E}$ diagram.
The semi-circles correspond to X-ray (orange) and TeV (black) PWNe, their sizes are proportional to logarithms 
of the corresponding PWN luminosities. The small black dots denote the pulsars from the ATNF catalog 
\cite{2005AJ....129.1993M}. Pulsars with PWNe detected by {\sl Fermi} are marked by stars. 
{\em Top right:} TeV-to-X-ray luminosity ratio vs. pulsar's age for PWNe and PWN candidates. 
Limits are shown by arrows. The PWNe detected by {\sl Fermi} are marked by stars.
Uncertain detections are shown by thin lines. The dotted lines corresponds to $L_\gamma = L_X$. 
{\em Bottom left:} TeV vs.\ X-ray photon indices for PWNe and PWN candidates. 
{\em Bottom right:} X-ray luminosities of PWNe and PWN candidates vs.\ pulsar's $\dot{E}$.
TeV PWNe and TeV PWN candidates are shown with grey thick error bars.
The dotted straight lines correspond to constant X-ray efficiencies; the upper bound, $\log L_X^{\rm cr} =
1.51logdot{E} -21.4$, is shown by a dashed line. The PWNe detected in GeV by {\sl Fermi} are marked by stars.
Figures adapted from \cite{olegbook2}. In all panels PWN of PSR J2032+4127 is marked in red.
}
\label{fig:pwne}
\end{figure}

The majority of identified Galactic TeV gamma-ray sources are pulsar wind nebulae (PWNe)
(\citet{2012ASPC..466..167K}).  Adopting the PWN scenario for \vtev\ with PSR J2023+4157 as the pulsar 
powering the TeV PWN, one can compare this source to other X-ray/TeV PWNe and PWNe candidates.  PSR 
J2032+4127 is one of the oldest and weakest pulsars (in terms of $\dot{E}$) whose PWN is detected both in 
X-ray and TeV gamma-rays, (see Figure \ref{fig:pwne}), Geminga being the other notable example. The TeV 
gamma-ray spectrum of \vtev\ is fit by a power-law with spectral index $\simeq2$, one of the hardest among PWNe 
and PWNe candidates. Because of this hard index, the spectrum must exhibit a cut-off not too far from 10 TeV 
due to the Klein-Nishina effect in order 
to be consistent with the PWN interpretation. The X-ray luminosity of the 
PWN (at d=1.7 kpc) is unremarkable (see Figure \ref{fig:pwne}) and consistent with the spin-down 
properties of the pulsar. Finally, the distance-independent ratio of the TeV to X-ray luminosity for 
\vtev\ is fairly well constrained and consistent with the expectations.

Interestingly, there is another TeV gamma-ray source (HESS J1646-458; \citet{2012A&A...537A.114A}; potentially a PWN of PSR J1648-4611)
whose spin-down properties are not too different from those of PSR~J2032$+$4127 apparently also located within the 
molecular cavity in the general direction of Westerlund 1 \citep{2010ApJ...713L..45L}. However, the X-ray 
PWN of PSR J1648-4611 has not yet been detected and its TeV PWN classification still remains to be proved.

\section{Conclusions}

VERITAS has made a deep observation of \tev\ resulting in a significance of 8.7 standard deviation
which is currently the most sensitive measurement made on this source in gamma rays
from 0.5 to 20 TeV. The position and the measured energy spectrum are found to 
be consistent with previous measurements made by other IACTs. 
The latter is fit well by a power-law with no evidence of a cutoff. 
The centroid of the emission has been measured to a greater accuracy than before, and for the first time
an intrinsic asymmetry in the morphology has been found. 
Almost all the TeV gamma-ray emission has been found to come from a region 
that is seen as a void in both radio and infrared wavebands. 
After considering multiwavelength data, we favor a relic pulsar wind nebulae scenario for \tev\ powered by 
the pulsar PSR J2032+4157.
However the possibility that the TeV gamma rays are produced by stellar winds cannot be ruled out, despite 
the relatively fewer number of massive stars in the void.

\section*{Acknowledgments} 

This research is supported by grants from the U.S. Department of Energy Office of Science, the U.S. National 
Science Foundation and the Smithsonian Institution, by NSERC in Canada, by Science Foundation Ireland (SFI 10/RFP/AST2748) and by STFC in the 
U.K. We acknowledge the excellent work of the technical support staff at the Fred Lawrence Whipple Observatory and at the collaborating 
institutions in the construction and operation of the instrument. 
GH acknowledges support through the Young Investigators Program of the Helmholtz Association.
The work by OK was supported by NASA grants NNX09AC84G and NNX09AC81G. 

\bibliography{tev2032}

\end{document}